\documentclass{ws-procs975x65}
\def\beq{\begin{equation}}
\def\eeq{\end{equation}}
\begin{document}

\title{Gravitational Waves From Pulsars Due To Their Magnetic Ellipticity} 
%/ Magnetic ellipticity of pulsars and gravitational wave emission / Magnetic ellipticity and gravitational waves from pulsars }
\author{Jos\'e C. N. de Araujo$^*$ }
\address{Divis\~ao de Astrof\'isica, Instituto Nacional de Pesquisas Espaciais,\\
S. J. Campos, SP 1227-010, Brazil\\
$^*$E-mail: jcarlos.dearaujo@inpe.br}

\author{Jaziel G. Coelho$^{**}$} 
\address{Departamento de F\'isica, Universidade Tecnol\'ogica Federal do Paran\'a\\
Medianeira, PR 85884-000, Brazil\\
$^{**}$E-mail: jazielcoelho@utfpr.edu.br}

\author{Samantha M. Ladislau$^\dagger$ and C\'esar A. Costa$^\ddagger$}
\address{Divis\~ao de Astrof\'isica, Instituto Nacional de Pesquisas Espaciais,\\
S. J. Campos, SP 1227-010, Brazil\\
$^\dagger$E-mail: samantha.ladislau@inpe.br\\
$^\ddagger$E-mail: cesar.costa@inpe.br}

\begin{abstract}
We discuss some aspects of de Araujo, Coelho and Costa~\cite{arau16,arau17} concerning the role of a time dependent magnetic ellipticity on the pulsars' braking indices and on the putative gravitational waves (GWs) these objects can emit. Since only nine of more than 2000 known pulsars have accurately measured braking indices, it is of interest to extend this study to all known pulsars, in particular as regards GW generation. 
In summary, our results show a pessimistic prospect for the detection of GWs generated by these pulsars, even for advanced detectors such as aLIGO and AdVirgo, and the planned Einstein Telescope, would not be able to detect these pulsar, if the ellipticity has magnetic origin.
\end{abstract}

%\keywords{Sample file; \LaTeX; MG14 Proceedings; World Scientific Publishing.}
\keywords{Pulsars; Gravitational Waves; braking index}

\bodymatter

%%%%%%%%%%%%%%%%% now a standard article style for the most part

\section{Ellipticity of Magnetic Origin and Gravitational Waves from Pulsars}

If the magnetic field and (or) the angle between the axes of rotation and the magnetic dipole of the pulsars are independent of time, the combination of magnetic dipole and gravitational wave (GW) brakes could only explain braking index (n) in the interval $3 < n < 5$. The observations, however, show that only PSR J1640-4631 has braking index in this interval, as can be seen in Table \ref{ta1}. 
In particular, we consider this issue in the context of magnetic ellipticity~\cite{arau16}. It is worth  stressing that the magnetic field and the angle between the axes of rotation and the magnetic dipole of the pulsars are dependent on time.

Recall that the equatorial ellipticity is given by          
\begin{equation}
\epsilon=\frac{I_{xx}-I_{yy}}{I_{zz}},
\end{equation}
where $I_{xx}$, $I_{yy}$, $I_{zz}$  are the moment of inertia with respect to the rotation axis, $z$, and along directions perpendicular to it.

The pulsar is deformed by its own dipole magnetic field. Such deformation associated with the fact that the axes of rotation and of the magnetic dipole are misaligned generates an ellipticity given by (see, e.g., Bonazzola and Gourgoulhon~\cite{bona96}; Konno et al~\cite{konn00}; de Freitas Pacheco and Regimbau~\cite{regi06}):
\begin{equation}
\epsilon_B = \kappa\frac{B_0^2 R^4}{G M^2}\sin^2\phi, \label{eq:epsilonB}
\end{equation}

where $B_0$ is the dipole magnetic field, $R$ and $M$ are the radius and the mass of the star respectively, $\phi$ is the angle between the rotation and magnetic dipole axes, whereas $\kappa$ is the distortion parameter, which depends on both the star equation of state (EoS) and the magnetic field configuration~\cite{regi06}. 
 We consider that  $\kappa = 10 - 1000$,  as suggested by numerical simulations~\cite{bona96, regi06}.

\begin{table}
\tbl{The periods ($P$) and their first derivatives ($\dot P$) for pulsars with known braking indices ($n$) (see also ATNF catalog~\citep{atnf03,manc05}).}
{\begin{tabular}{@{}llrrrr@{}} 
\toprule
 & Pulsar & $P$~(s) &$\dot{P}~(10^{-13}$~s/s) &n$^\diamond $ & \\ 
\colrule
& PSR J1734-3333      &1.17  &22.8 &$0.9\pm0.2$~\cite{espi11}     & \\
& PSR B0833-45 (Vela) &0.089 &1.25 &$1.4\pm0.2$~\cite{lyne96}     & \\
& PSR J1833-1034      &0.062 &2.02 &$1.8569\pm0.0006$~\cite{roy12}& \\
& PSR J0540-6919      &0.050 &4.79 &$2.140\pm0.009$~\cite{livi07} & \\ %(B0540-69)
& PSR J1846-0258      &0.324 &71   &$2.19\pm0.03$~\cite{arch15}   & \\
& PSR B0531+21 (Crab) &0.033 &4.21 &$2.51\pm0.01$~\cite{lyne93}   & \\
& PSR J1119-6127      &0.408 &40.2 &$2.684\pm0.002$~\cite{welt11} & \\
& PSR J1513-5908      &0.151 &15.3 &$2.839\pm0.001$~\cite{livi07} &  \\ % (B1509-58)
& PSR J1640-4631      &0.207 &9.72 &$3.15\pm0.03$~\cite{arch16}   & \\
\botrule
\end{tabular}}
\begin{tabnote}$^\diamond n \equiv f_{\rm rot}\,{\ddot f}_{\rm rot}/{\dot{f}^2_{\rm rot}}$, where $f_{\rm rot} = 1/P$ is the rotating frequency, 
$\dot{f}_{\rm rot}$ and $\ddot{f}_{\rm rot}$ are their time derivatives.\\\end{tabnote}
\label{ta1}
\end{table}

Recall that the power emitted by a rotating magnetic dipole is given by~\cite{padm01}
\begin{equation}
\dot{E}_{\rm d}= -\frac{16\pi^4}{3}\frac{B_0^2 R^6\sin^2\phi}{c^3}f_{\rm rot}^4,  \label{Ed}
\end{equation}
\noindent  and the power loss via GW emission reads~\cite{shap83}
\begin{equation}
\dot{E}_{\rm GW} = -\frac{2048\pi^6}{5}\frac{G}{c^5}I^2\epsilon^2 f_{\rm rot}^6. \label{EGW}
\end{equation}

Also, the total energy of the pulsar is provided by its rotational energy, $E_{\rm rot} = 2  \pi^2If_{\rm rot}^2$, and any change on it is 
given by $\dot{E}_{\rm d}$ and $\dot{E}_{\rm GW}$, namely
\begin{equation}
\dot{E}_{\rm rot}\equiv \dot{E}_{\rm GW} +\dot{E}_{\rm d} \label{Erotdef}.
\end{equation}

Now, from the definition of the braking index (see, e.g., the note in Table \ref{ta1}), one can easily obtain 
that\footnote{The detailed derivation of Eq.~\ref{neta} can be found in de Araujo, Coelho \& Costa~\cite{arau16}.} 
\begin{equation}
n=3+2\eta-2\frac{P}{\dot P}\left(1+\eta\right)\left[\frac{\dot B_0}{B_0}+\dot{\phi}\cot{\phi}  \right], \label{neta}
\end{equation}
\noindent where $\eta$ is defined in such a way that $\dot{E}_{\rm GW} = \eta \dot{E}_{\rm rot}$, which 
is interpreted as the efficiency of GW generation.  In de Araujo, Coelho \& Costa~\cite{arau16} it is also shown that with Eq.~\ref{neta}
one can explain, in principle, the braking indices of the pulsars of Table \ref{ta1}.

Recall that the GW amplitude generated by a pulsar reads
\begin{equation}
h^2 = \frac{5}{2}\frac{G}{c^3}\frac{I}{r^2}\frac{\mid\dot{f}_{\rm rot}\mid}{f_{\rm rot}}.
\end{equation}
\noindent This equation considers that the spindown is due to gravitational waves only, i.e., n = 5 (spindown limit - SD).  

From the definition of $\eta$ one obtains that $\dot{\bar{f}}_{\rm rot} = \eta \dot{f}_{\rm rot}$, i.e., the part of 
the spindown related to the GW emission brake. Thus, one can obtain an equation for the GW amplitude that holds for n $ < $ 5, namely
\begin{equation}
\bar{h}^2 = \frac{5}{2}\frac{G}{c^3}\frac{I}{r^2}\frac{\mid\dot{\bar{f}}_{rot}\mid}{f_{rot}} =  \frac{5}{2}\frac{G}{c^3}\frac{I}{r^2}\frac{\mid\dot{f}_{\rm rot}\mid}{f_{\rm rot}} \, \eta . \label{heta}
\end{equation}

Recall that the GW amplitude also reads
\begin{equation}
h = \frac{16\pi^2G}{c^4} \frac{I\epsilon f_{\rm rot}^2}{r},
\end{equation}
\noindent (see, e.g, Shapiro and  Teukolsky\cite{shap83}). Combining both equations for the GW amplitude one obtains
\begin{equation}
\epsilon = \sqrt{\frac{5}{512\pi^4} \frac{c^5}{G}\frac{\dot{P}P^3}{I}\eta}. \label{epet}
\end{equation} 

Now, for a purely magnetic brake we have 
\begin{equation}
\bar{B}_0\sin^2\phi = \frac{3 I c^3}{4 \pi^2 R^6} P \dot{P},
\end{equation}
where $\bar{B}_0$ would be the magnetic field whether the break were purely magnetic. If there is also a GW brake contribution we have that $B_0 < \bar{B}_0$. 
Combining the definition of  $\eta$ and Eq. \ref{epet} one obtains after some algebraic manipulation the following equation for the efficiency $\eta$ 
\begin{equation}
\eta = 1 - \left(\frac{B_0}{\bar{B}_0} \right)^2,
\end{equation}
which is obviously lower than one, as it should be. Substituting this last equation into Eq.\ref{eq:epsilonB} we obtain
\begin{equation}
\epsilon = \frac{3Ic^3}{4\pi^2GM^2R^2}P\dot{P} \left(1 - \eta\right) \kappa. \label{eek}
\end{equation}
Finally, substituting this last equation into equation \ref{epet}, we obtain
\begin{equation}
\eta = \frac{288}{5}\frac{I^3c}{GM^4R^4}\frac{\dot{P}}{P}\left( 1-\eta \right)^2 \kappa^2. \label{ek}
\end{equation}

Notice that with Eqs.\ref{eek} and \ref{ek} we obtain $\epsilon$ and $\eta$ in terms of $M$, $R$, $I$, $P$ and $\dot{P}$ for a given value of $\kappa$. 
Since in practice $\eta \ll 1$, the following useful equations are obtained  
\begin{equation}
\epsilon \simeq \frac{3Ic^3}{4\pi^2GM^2R^2}P\dot{P}\kappa \label{eeka}
\end{equation}
and 
\begin{equation}
\eta \simeq \frac{288}{5}\frac{I^3c}{GM^4R^4}\frac{\dot{P}}{P} \kappa^2.  \label{eka}
\end{equation}

We now calculate $\epsilon_{B}$ and $\eta$ for the pulsars of Table \ref{ta1}. We then adopt fiducial values for M, R and I. We adopt $\kappa = 10$ and 1000, which have the same orders of magnitude of the values considered by, e.g., Regimbau and  de  Freitas  Pacheco\cite{regi06} . 

In Table \ref{ta2} we present the result of these calculations. Even for the extremely optimistic case, the value of the ellipticity is at best $\epsilon_{B} \sim 10^{-5}$ (for \mbox{PSR J1846-0258}) and the corresponding efficiency $\eta \sim 10^{-8}$. Therefore, the amplitude of the GW in this case would be four orders of magnitude lower than the spindown limit ($\eta=1$). Thus, even advanced detectors such as aLIGO and AdVirgo, and the planned Einstein Telescope, would not be able to detect these pulsars.

\begin{table}
\tbl{$\epsilon$ and $\eta$ for $\kappa = 10\; (1000)$ for the Pulsars of Table \ref{ta1}.}
{\begin{tabular}{@{}llllr@{}} 
\toprule
& Pulsar &  $\epsilon$ & $\eta$  & \\ 
\colrule
& PSR J1734-3333      & $1.2\times 10^{-7(-5)}  $  & $1.1\times 10^{-13(-9)} $ &  \\
& PSR B0833-45 (Vela) & $4.9\times 10^{-10(-8)} $  & $8.3\times 10^{-14(-10)}$ &  \\
& PSR J1833-1034      & $5.5\times 10^{-10(-8)} $  & $1.9\times 10^{-13(-9)} $ &  \\
& PSR J0540-6919      & $1.1\times 10^{-9(-7)}  $  & $5.7\times 10^{-13(-9)} $ &  \\ 
& PSR J1846-0258      & $1.0\times 10^{-7(-5)}  $  & $1.3\times 10^{-12(-8)} $ &  \\
& PSR B0531+21 (Crab) & $6.1\times 10^{-10(-8)} $  & $7.5\times 10^{-13(-9)} $ &  \\
& PSR J1119-6127      & $7.2\times 10^{-8(-6)}  $  & $5.8\times 10^{-13(-9)} $ &  \\
& PSR J1513-5908      & $1.0\times 10^{-8(-6)}  $  & $6.0\times 10^{-13(-9)} $ &  \\ 
& PSR J1640-4631      & $8.9\times 10^{-9(-7)}  $  & $2.8\times 10^{-13(-9)} $ &  \\
\botrule
\end{tabular}}
\label{ta2}
\end{table}

\begin{figure}[h]
\begin{center}
\includegraphics[width=8cm]{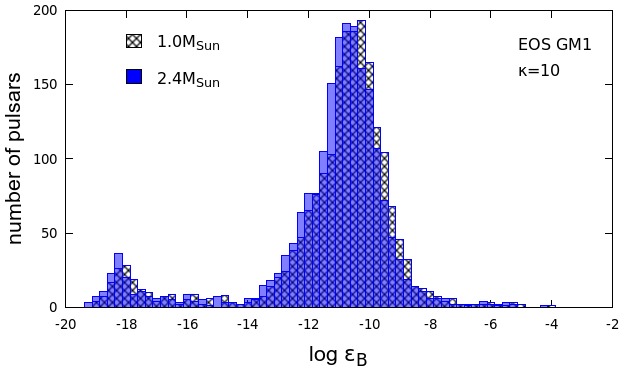}
\end{center}
\caption{Ellipticity histogram for the pulsars of ATNF Catalog for $\kappa = 10$.}
\label{fig1}
\end{figure}

Notice that Eqs. \ref{eeka} and \ref{eka} do not depend on the braking index n. Consequently, we can calculate such quantities 
for the pulsars of the ATNF Pulsar Catalog. We refer the reader to the paper by de Araujo, Coelho and Costa\cite{arau17} 
for details .
In Fig. \ref{fig1}  we show an interesting histogram with the data of the ATNF Catalog, namely, the number of pulsars for $\log \epsilon_{B}$ bin. 
Note the high number of pulsars concentrated around $\sim 10^{-10}\, (10^{-8})$  for $ k = 10 \,(1000)$.  The values of $\eta$ are also extremely
small, a histogram can be found in de Araujo, Coelho \& Costa\cite{arau17}, where can be seen a peak at $10^{-16} - 10^{-15}$.  

\begin{figure}[h]
\begin{center}
\includegraphics[width=8cm]{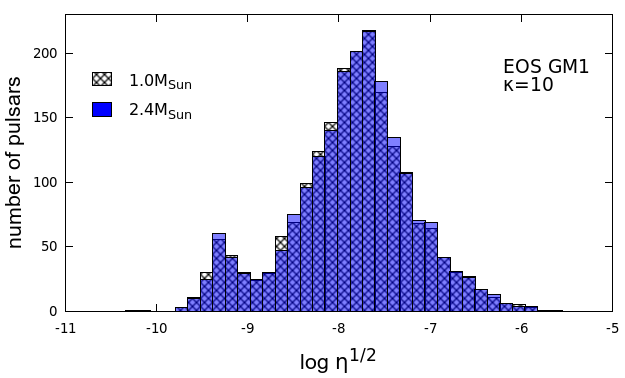}
\end{center}
\caption{Histogram of $\eta^{1/2} = h/h^{SD}$  (spin-down ratio) for the pulsars of ATNF Catalog for $\kappa = 10$.}
\label{fig2}
\end{figure}

These extremely small values of $\epsilon_{B}$ and $\eta$ imply that the GW amplitudes are at best seven orders of magnitude 
smaller than those obtained by assuming the spindown limit (SD), being therefore hardly detected (see Fig.~\ref{fig2}). 

\section{Final Remarks}

We present an expression for the braking index considering that the ellipticity is of magnetic dipole origin and time dependent. In this context, we model 
the braking indices of the 9 pulsars that have such measured quantities accurately. Then we calculate the amplitudes of the GWs generated by these 9 pulsars. Summing up, we conclude that these amplitudes are too small to be detected. For example, the pulsar PSR J1846-0258 would need to be observed for over 1000 years to 
be detected by the Einstein Telescope.

Since the equations for $\eta$, $\epsilon_{B}$  and  $h$ are independent of n, we extend our study for most of the pulsars of the "ATNF Pulsar Catalog". 
Regarding detectability, the prospects remain pessimistic, since the ellipticity generated by the magnetic dipole is extremely small, the corresponding amplitude of GWs is much smaller than the amplitude obtained via the spindown limit.

\section*{Acknowledgments}
J.C.N.A thanks FAPESP (2013/26258-4) and CNPq (307217/2016-7) for partial support. J.G.C. is likewise grateful to the support of CNPq (421265/2018-3 and 305369/2018-0). S.M.L. and C.A.C. acknowledge CAPES for financial support.

%\bibliographystyle{ws-procs975x65}       
%\bibliography{ref.bib} 

\begin{thebibliography}{00}
\bibitem{arau16} J.C.N.de Araujo, J.G. Coelho and C.A. Costa, {\em ApJ} {\bf 831} 35 (2016)
\bibitem{arau17} J.C.N.de Araujo, J.G. Coelho and C.A. Costa, {\em EPJC} {\bf 77}, 350 (2017)
\bibitem{bona96} S. Bonazzola and E.  Gourgoulhon, {\em A\&A}  {\bf 312}, 675 (1996)
\bibitem{konn00} K. Konno, T. Obata and Y. Kojima, {\em A\&A} {\bf 356}, 234 (2000)
\bibitem{regi06} T. Regimbau and J.A.  de  Freitas  Pacheco,  {\em A\&A} {\bf 447}, 1 (2006)
\bibitem{atnf03} CSIRO.  ATNF Pulsar Catalogue. \\ \url{http://www.atnf.csiro.au/people/pulsar/psrcat/} (2003).
\bibitem{manc05} R.N. Manchester, G.B. Hobbs, A. Teoh and M. Hobbs, {\em The Astronomical Journal} {\bf 129}, 1993 (2005).
\bibitem{espi11} C.M. Espinoza, A.G. Lyne, M. Kramer, R.N. Manchester and V.M. Kaspi, {\em ApJL} {\bf 741}, L13 (2011).
\bibitem{lyne96} A.G. Lyne, R.S. Pritchard, F. Graham-Smith and F. Camilo, {\em Nature} {\bf 381}, 497 (1996).
\bibitem{roy12}  J. Roy, Y. Gupta and W. Lewandowski, {\em MNRAS} {\bf 424}, 2213 (2012).
\bibitem{livi07} M.A. Livingstone, V.M. Kaspi, F.P. Gavriil, R.N. Manchester, E.V.G. Gotthelf and L. Kuiper, {\em Astrophys Space Sci} {\bf 308}, 317 (2007).
\bibitem{arch15} R.F. Archibald, V.M. Kaspi, A.P. Beardmore, N. Gehrels and J.A. Kennea, {\em ApJ} {\bf 810}, 67 (2015). 
\bibitem{lyne93} A.G. Lyne, R.S. Pritchard and F. Graham-Smith, {\em MNRAS} {\bf 265}, 1003 (1993)
\bibitem{welt11} P. Weltevrede, S. Johnston and C.M. Espinoza, {\em MNRAS} {\bf 411}, 1917 (2011). 
\bibitem{arch16} R.F. Archibald, E.V. Gotthelf, R.D. Ferdman, V.M. Kaspi, S. Guillot, F.A. Harrison, E.F. Keane, M.J. Pivovaroff, D. Stern, S.P. Tendulkar and J.A. Tomsick, {\em ApJl} {\bf 819}, L16 (2016).   
\bibitem{padm01} T. Padmanabhan, in {\it Theoretical Astrophysics - Volume 2, Stars and Stellar Systems} (Cambridge University Press, 2001).
\bibitem{shap83} S.L. Shapiro and S.A. Teukolsky, in {\it Black Holes, White Dwarfs, and Neutron Stars: The Physics of Compact  Objects}  (New York, Wiley-Interscience, 1983)

	
\end{thebibliography}

\end{document}